\documentclass[12pt]{article}
\usepackage{amsmath}
\usepackage{amsfonts}
\usepackage{amssymb}
\usepackage{amsthm}
\usepackage{amscd}

\def\be{\begin{equation}}
\def\ee{\end{equation}}
\def\bea{\begin{eqnarray}}
\def\eea{\end{eqnarray}}

\def\p{\partial}
\def\a{\alpha}

\def\d{\delta}

\def\k{\kappa}

\def\vfi{\varphi}
\def\T{\Theta}

\def\ra{\rangle}
\def\la{\langle}

\def\wt{\widetilde}

\def\erf{{\rm erf}\,}

\newcommand{\Sc}{Schr\"odinger }

\newtheorem{Th}{Theorem}
\newtheorem{Rem}{Remark}

\begin{document}

\title {\large\bf Exact propagators for complex SUSY partners of real potentials}

\author{\normalsize
 Boris F. Samsonov and Andrey M. Pupasov
}

\date{\small Department of Physics, Tomsk State
 University, 36 Lenin Avenue, 634050 Tomsk, Russia}

\maketitle

\begin{abstract}
A method for calculating
exact propagators for those complex potentials
with a real spectrum
which are SUSY
partners of real potentials is presented.
It is illustrated by examples of propagators for
some complex SUSY partners of the harmonic oscillator
and zero potentials.
\end{abstract}


Recently a considerable attention has been paid to investigating
different properties of non-Hermitian Hamiltonians (see
 e.g. \cite{ChJP}).
One of the reasons of that is an
 attempt to generalize the quantum mechanics by accepting
 non-Hermitian Hamiltonians with purely real spectrum (see e.g.
 \cite{ChJP,Bender} and references therein).
The propagator being the coordinate representation
of the evolution operator
 is one of the important objects in quantum mechanics since
 it permits to describe the evolution of a quantum system
 for an arbitrary initial state.
 On the
 other hand the method of supersymmetric quantum mechanics (SUSY
 QM) is one of the main methods for getting complex exactly
 solvable Hamiltonians \cite{Cannata,0602101}.
 Moreover, because of its
 nice property to convert a non-diagonalizable Hamiltonian into
 diagonalizable forms and delete spectral singularities from the
 continuous spectrum of a non-Hermitian Hamiltonian
\cite{My} it was presumed \cite{SaSh} that SUSY QM may become an
essential ingredient of the infant complex quantum mechanics.
The aim of this Letter is to show that SUSY QM may be useful for
finding propagators for those complex
potentials which are
SUSY partners of real potentials for which
both the propagator and the Green's function are known thus giving
an additional argument in favor of the above thesis.
For simplicity we will consider only time-independent potentials
although corresponding time-dependent technique is also available
\cite{tBS}.

We consider the one-dimensional \Sc equation with a
complex-valued
potential $V_c(x)$
\be\label{ste}
[\,i \frac{\p}{\p t}-h_c\,]\,\Phi(x,t)=0\,,
\ee
\be
h_c=-\frac{\p^2}{\p x^2}+V_c(x)\,,\quad
x\in\Bbb R\,.
\ee
Since the potential $V_c(x)$ is time independent
(stationary) solutions to equation (\ref{ste}) are expressed in terms of
solutions of the stationary equation
\be\label{shc}
h_c\vfi_E(x)=E\vfi_E(x)
\ee
in the usual way, $\Phi(x,t)=e^{-i Et}\vfi_E(x)$.
But if initially the quantum system is prepared in the state
$\vfi_0(x)$ which is not a state with the definite value of the
energy we need to know the propagator for describing the evolution
of a quantum system.
We use the usual definition of the
propagator $K_c(x,y,t)$ as a solution to equation (\ref{ste}) with
respect to variables $x$ and $t$
with the delta-like initial condition
\be\label{Kc}
K_c(x,y;0)=\delta(x-y)\,.
\ee
If $K_c(x,y,t)$ is given the function
\be\label{Phi}
\Phi(x,t)=\int_{-\infty}^{\infty}K_c(x,y;t)\vfi_0(y)dy
\ee
is a solution to equation (\ref{ste}) with the initial condition
$\Phi(x,0)=\vfi_0(x)$.

First of all we note that to be able to associate the function
$\Phi(x,t)$ with a state of a quantum system the integral in
(\ref{Phi}) should converge and both the function $\vfi_0(x)$ and
$\Phi(x,t)$ should belong to a certain class of functions.
To present our method in its simplest form we will
make several assumptions which shall simplify essentially our
presentation keeping at the same time the essence of the method.
Since in (\ref{Phi}) the usual (Lebesgue) integration is involved
it is natural to suppose that both $\vfi_0(x)$ and $\Phi(x,t)$ are
square integrable.
This means that the Hilbert space, where the
operator (non-Hermitian Hamiltonian)
$h_c$ associated with the differential expression
$-\p^2 x/\p x^2+V_c(x)$ 'lives', is the usual space
$L_2(\Bbb R)$ and the equation (\ref{shc}) creates an eigenvalue
problem for $h_c$ which is defined on a dense
domain from $L_2(\Bbb R)$.
Similar eigenvalue problems were under an
intensive study by mathematicians in the Soviet Union
 in the period
between 50th and 70th of the previous century.
Results of these investigations are mainly summarized in
books \cite{Krein,Naimark} to which we refer the interested reader
where he can, in particular, find the strict definition of the
spectrum, eigenfunctions, associated functions, domains of
definition of operators created by non-Hermitian
differential expressions
and many other properties of differential equations and
related
 non-selfadjoint operators.
Here
we would like to mention that the first essential result in
this field was obtained by Keldysh \cite{Keldysh} who proved the
completeness of the set of eigenfunctions and associated functions
for a non-selfadjoint operator
and results by Lidskiy \cite{Lid}    having the direct relation to
 the current paper. In particular, Lidskiy made a
deep analysis of conditions on the potential $V_c$ leading to an
operator $h_c$ which is uniquely defined by its closure and has a
purely discrete spectrum with a complete set of eigenfunctions and
associated functions.

Our next essential assumptions is that $h_c$ has a purely discrete
 spectrum, its set of associated functions is empty,
it is diagonalizable, and its set of eigenfunctions
$\phi_n(x)$, $n=0,1,\ldots$
is
complete in the space $L_2(\Bbb R)$.
If $h_c^+$ is the adjoint differential expression it creates in
the Hilbert space the adjoint operator with the eigenfunctions
$\wt\phi_k(x)$ which also form a complete set in $L_2(\Bbb R)$.
Moreover, if $E_n$ is an eigenvalue of $h_c$ then $E_n^*$
 (asterisk means the complex conjugation) is an
eigenvalue of $h_c^+$ so that
$h_c^+\wt\phi_n=E_n^*\wt\phi_n$.
Note that neither $\{\phi_n\}$ nor $\{\wt\phi_n\}$, $n=0,1,\ldots$
form orthogonal systems but
functions
$\wt\phi_k$ are biorthogonal with $\phi_n$
and they can always be normalized such that
(see e.g. \cite{Naimark})
\be\label{biortn}
\int_{-\infty}^\infty\wt\phi_k^*(x)\phi_n(x)dx=\d_{nk}\,.
\ee
The completeness of the set of eigenfunctions of $h_c$ means that
any $\phi\in L_2(\Bbb R)$ can be developed
into the Fourier series over the set $\{\phi_n\}$,
$
\phi(x)=\sum_{n=0}^\infty c_n\phi_n(x).
$
Using the biorthonormality relation (\ref{biortn}) we can find the
coefficients $c_n$ in the usual way and put them back into the
same relation thus obtaining the symbolical form of the
completeness condition of the set of eigenfunctions of $h_c$
\be\label{compl}
\sum_{n=0}^\infty\wt\phi_n^*(x)\phi_n(y)=\delta(x-y)\,.
\ee

Next we are  assuming  that the spectrum $h_c$ is
 real. Therefore the adjoint eigenvalue problem coincides with the
 complex conjugate form of equation (\ref{shc}) so that
 $\wt\phi_n(x)=\phi_n^*(x)$.
Under these assumptions equations (\ref{biortn}) and (\ref{compl})
become (cf. with \cite{SRoy})
\be\label{ort}
\int_{-\infty}^{\infty}\phi_n(x)\phi_k(x)dx=\delta_{nk}\,,
\ee
\be\label{poln1}
\sum\limits_{n=0}^{\infty}\phi_n(x)\phi_n(y)=\delta(x-y)\,.
\ee
From here it follows the Fourier series expansion of the
propagator in terms of the basis functions $\phi_n$:
\be\label{decpr}
K_c(x,y;t)=\sum_{n=0}^\infty \phi_n(x)\phi_n(y)e^{-iE_n t}\,.
\ee
Indeed, just like in the conventional Hermitian case the initial
condition $\Phi(x,0)=\vfi_0(x)$ for function
(\ref{Phi}) with $K_c$ of form (\ref{decpr})
 follows from (\ref{poln1}) and the fact that $\Phi(x,t)$
 (\ref{Phi}) satisfies equation (\ref{ste}) follows from
 the property of the functions $\phi_n$ to be eigenfunctions of $h_c$
 with the eigenvalues $E_n$.

Especial role
 between all non-selfadjoint operators is played by
pseudo-Hermitian operators
first introduced by Dirac and Pauli and latter used by Lee and
Wick \cite{DPLW} to overcome some difficulties related with using
Hilbert spaces with an indefinite metric
and their recent generalization (weak pseudo-Hermiticity)
by Solombrino and Scolarici \cite{SS}
since there are strict indications that these operators are
 the most appropriate candidates for replacing
selfadjoint operators while generalizing the conventional
quantum mechanics  by accepting non-Hermitian operators
\cite{SS,Mosta}.

 Another useful discussion is that the form (\ref{decpr}) for the
 propagator may be interpreted as the coordinate representation
 of the abstract evolution operator.\footnote{
This property has been demonstrated in the report of the
anonymous referee.}
 To show this we introduce ket-vectors (kets) $|\phi_n\ra$ as
 eigenvectors of $h_c$ and bra-vectors (bras) $\la\wt\phi_n|$
 as functionals acting in the space of kets according
 to\footnote{Without going into details we note that using the
 system
 $|\phi_n\rangle$ one can construct the Hilbert space $H$ so that
 the set of all finite linear combinations of $\phi_n$ is dense in
 $H$ and formula (\ref{biort}) uniquely defines a functional
 in $H$, see e.g. \cite{Nikol}.}
\be\label{biort}
 \la\wt\phi_n|\phi_k\ra=\d_{nk}\,.
 \ee
  Kets corresponding to
 the previous bras are just properly normalized eigenvectors of
 $h_c^+$
 which is
 defined by the adjoint eigenvalue problem where $V_c(x)$ is
 replaced by its complex conjugate $V_c^*(x)$ so that
 (\ref{biort}) is nothing but the same biorthogonality condition
 (\ref{ort}) written in the abstract representation.
 As usual the coordinate representation of the above abstract
 eigenvectors are
 $\phi_n(x)=\la x|\phi_n\ra=\la\wt\phi_n|x\ra$
 where $|x\ra$ is an eigenvector of the coordinate operator.
 The completeness condition in the abstract form now reads
 \be
\sum_{n=0}^\infty|\phi_n\ra\la\wt\phi_n|=1
 \ee
and the formula
 \be
h_c=\sum_{n=0}^\infty|\phi_n\ra E_n\la\wt\phi_n|
 \ee
presents the spectral decomposition of the Hamiltonian $h_c$.
Now in the known way \cite{SS,Mosta} one can introduce an
automorphism $\eta$ to establish the property that $h_c$ is
(weakly) pseudo-Hermitian and construct a basis in which $h_c$
takes a real form.
We will not go into further details of the known properties of
(weakly)
pseudo-Hermitian operators since this is not the aim of this
paper. The interested reader can consult papers
\cite{DPLW,SS,Mosta} and the recent preprint \cite{Mis} where
biorthogonal systems are widely used in the study of different
properties of non-Hermitian Hamiltonians.
Our last comment here is that the abstract evolution operator
given by its spectral decomposition
\be\label{Ut}
U(t)=\sum_{n=0}^\infty|\phi_n\ra e^{-iE_nt}\la\wt\phi_n|
\ee
written in the coordinate representation
$K_c(x,y,t)=\la x|U(t)|y\ra$ is just the propagator
(\ref{decpr}).

We would like to emphasis that conditions
 (\ref{ort}) and (\ref{poln1})
 have almost the usual form, only
the complex conjugation is absent.
Therefore they coincide with the corresponding equations for the
Hermitian Hamiltonians in case when their eigenfunctions are real.

The final assumption we make is that the Hamiltonian $h_c$ is a
SUSY partner of a Hermitian Hamiltonian $h_0$
with a purely discrete spectrum and a complete set of
eigenfunctions $\psi_n$ which always can be chosen real
\[
 h_0=-\frac{\p^2}{\p x^2}+V_0(x)\,,
\quad h_0\psi_n(x)=E_n\psi_n(x)\,,\quad \psi_n(x)=\psi_n^*(x)\,,
\quad n=0,1,\ldots
\]
so that both the completeness and normalization conditions are given
by equations (\ref{poln1}) and (\ref{ort}) respectively with the
replacement $\phi_n\to\psi_n$.

According to the general scheme of SUSY QM (see e.g. \cite{BS})
operators $h_0$ and $h_c$ are (1-)SUSY partners if and only if
 there exists a first order differential operator $L$ such that
 \be\label{intert}
Lh_0=h_cL\,.
\ee
Operator $L$ of the form
\be\label{L}
L=-u'(x)/u(x)+\p/\p_x\,,
\ee
where
the prime means the derivative with respect to $x$
and the function $u(x)$ is a solution to equation
\be\label{Sa}
 h_0u(x)=\a u(x)\,,
\ee
exists for any $h_0$ and a rather restricted set of operators
$h_c$, the potentials $V_c$ of which are defined by
\[
 V_c=V_0-2[\log u(x)]''\,.
\]
The spectrum of $h_c$ may either (i) coincide with the spectrum of
$h_0$ or (ii) may differ from it by one (real) level which is absent
in the spectrum of $h_0$.
The case (i) may be realized only with a complex
parameter $\a$ which is called the factorization constant.
This
statement is a consequence of a proposition proved in
\cite{0602101}.
The case (ii) may be realized only for a real factorization
constant since $E=\a$ is just the discrete level of $h_c$ missing
in the spectrum of $h_0$ and we want that $h_c$ has a real spectrum.
Therefore in this case one has to choose
$u(x)$ as a linear combination of two real
linearly independent solutions to equation (\ref{Sa}).

Together with operator $L$ (\ref{L}) we need also its 'transposed
form' which we define as follows:
\be
L^t=-u'(x)/u(x)-\p/\p_x\,.
\ee
Just like in the usual SUSY QM the following factorizations take
place:
\be\label{fact}
L^tL=h_0-\a\,,\qquad LL^t=h_c-\a
\ee
which can easily be checked by the direct calculation.

One of the main features of the method is that for the most physically
interesting Hamiltonians $h_0$ operator (\ref{L})
has the property $L\psi_E(\pm\infty)=0$ provided
$\psi_E(\pm\infty)$.
As a result the set of functions
\be\label{phin}
\phi_n=N_nL\psi_n\,,\qquad n=0,1,\ldots
\ee
 is complete in
the space $L^2({\Bbb R})$ in case (i). In case (ii) we
have to add to this set the function $\phi_\a=N_\a/u$.
The normalization coefficients $N_n$ may be found by integration
by parts in equation (\ref{ort}) and with the help of
factorization property (\ref{fact}) which yields
\be\label{Nn}
 N_n=(E_n-\a)^{-1/2}\,.
 \ee

The main result of the present Letter is given by the following
\begin{Th}
The propagator  $K_c(x,y;t)$ of the \Sc equation with the
Hamiltonian $h_c$ related with $h_0$ by a SUSY transformation is
expressed in terms of the propagator
 $K_0(x,y;t)$
of the same equation with the Hamiltonian $h_0$ and the Green's
function  $G_0(x,y;E)$ of the stationary equation with the same
Hamiltonian as follows:\\
in case (i)
$K_c(x,y,t)=K_L(x,y,t)$\\
in case (ii)
$K_c(x,y,t)=K_L(x,y,t)+\phi_\a(x)\phi_\a(y)e^{-i\a t}$\\
where $K_L(x,y,t)$ is the 'transformed' propagator
\be\label{trpr1}
K_L(x,y,t)=L_xL_y\int_{-\infty}^{\infty}K_0(x,z,t)G_0(z,y,\a)dz\,.
\ee
Here $L_x$ is defined by (\ref{L}) and $L_y$ is the same operator
where $x$ is replaced by $y$.
\end{Th}

\begin{Rem}
We use the definition of the Green's function,
which we denote $G_0(x,y,E)$ where $E$ belongs
to the resolvent set of $h_0$,
as a coordinate
representation of the resolvent $(h_0-E)^{-1}$
(see e.g. \cite{MF}).
The Green's function is a kernel of an
integral operator acting in the space $L_2(\Bbb R)$.
There exists two equivalent representations for the Green's
function:\\
(a)
In terms of the basis functions $\psi_n(x)$
with eigenvalues $E_n$, $n=0,1,\ldots$
\be\label{G0}
G_0(x,y,E)=
\sum_{m=0}^{\infty}\frac{\psi_m(x)\psi_m(y)}{E_m-E}
\ee
provided $\psi_n(x)$ are real
and\\
(b) in terms of two special linearly independent solutions
to the
\Sc equation with the given value of $E$ (see an example below).
\end{Rem}

{\bf Proof}. We prove the theorem only for case (ii)
since case (i) does not have essentially new moments.

Using equations (\ref{decpr}), (\ref{phin}) and (\ref{Nn})
we get
\be\label{KcLL}
K_c(x,y,t)=L_xL_y
\sum_{m=0}^{\infty}\frac{\psi_m(x)\psi_m(y)}{E_m-\a}
e^{-iE_mt}+ \phi_\a(x)\phi_\a(y)e^{-i\a t}\,.
\ee
Because of the equality
\[
\psi_m(x)e^{-iE_mt}=\int_{-\infty}^{\infty}K_0(x,z,t)\psi_m(z)dz
\]
equation (\ref{KcLL}) assumes the form
\be\label{vc1}
K_c(x,y,t)=
L_xL_y
\int_{-\infty}^{\infty}K_0(x,z,t)
\sum_{m=0}^{\infty}\frac{\psi_m(z)\psi_m(y)}{E_m-\a}dz+
\phi_\a(x)\phi_\a(y)e^{-i\a t}\,.
\ee
Here under the summation sign we recognize the Green's function
(\ref{G0}) which ends the proof.\hfill $\square$

We have to note that both in case (i) and in case (ii) $\a$
belongs to the resolvent set of $h_0$. Hence, the Green's function
$G_0(x,y,\a)$ entering in (\ref{trpr1}) and (\ref{vc1})
is regular.
Another useful comment is that almost the same proof
is valid for the case when $h_0$ (and hence $h_c$) has a
continuous spectrum so that the theorem is valid for this case
also provided $h_c$ has no spectral singularities in the
continuous part of the spectrum.

To show how the theorem works we will consider a couple of typical
examples.

Consider first the harmonic oscillator Hamiltonian
$h_0(x)=-\p^2/\p x^2+x^2/4$
for which the propagator is well-known \cite{Feynman1}
\[
K_{0}(x,y,t)=
\frac{1}{\sqrt{4\pi i \sin t}}\,{\rm{e}}^{
\frac{i[(x^2+y^2)\cos t-2xy]}{4\sin t}}\,.
\]
To apply our theorem we need also the Green's function
of the oscillator Hamiltonian  at $E=\a$.
To find it we use the definition of the Green's function
in terms of two special solutions
$f_{l}(x,E)$ and $f_{r}(x,E)$
of the equation
\[
(h_0-E)f_{l,r}(x,E)=0\,, \qquad f_l(-\infty,E)=0\,,\quad
f_r(\infty,E)=0
\]
which is
\be\label{G-osc}
G(x,y,E)=[f_l(x,E)f_r(y,E)\T(y-x)+
f_l(y,E)f_r(x,E)\T(x-y)]/W(f_l,f_r)
\ee
where $\T$ is the Heaviside step function
and $W$ stands for the Wronskian.

In the simplest case we can choose $\a=-1/2$ and
\[
u(x)=e^{\frac{x^2}{4}}(C+\erf(x/\sqrt{2}))\,,
\quad {\rm Im}\,C\ne 0
\]
Functions $f_l$ and $f_r$ from (\ref{G-osc}) at $E=\a=-1/2$  read
\be\label{flr}
f_l(x,-1/2)=\sqrt{\pi/2}e^{\frac{x^2}{4}}(1+\erf(x/\sqrt{2}))\,,\quad
f_r(x,-1/2)=\sqrt{\pi/2}e^{\frac{x^2}{4}}(1-\erf(x/\sqrt{2}))\,.
\ee

The spectrum of the
complex-valued transformed potential
\be\label{pot1}
V_c(x)=\frac{x^2}{4}-1+2xQ_1^{-1}(x)e^{-\frac{x^2}{2}}+2Q_1^{-
2}e^{-x^2}\,,\quad
Q_1(x)=\sqrt{\frac{\pi}{2}}[C+\erf(x/\sqrt{2})]
\ee
consists of
 all oscillator energies $E_n=n+1/2$, $n=0,1,\ldots$
 and one additional level $E_{-1/2}=-1/2$ with the eigenfunction
 \be\label{psi12}
\phi_{-1/2}(x)=(2\pi)^{-1/4}\sqrt{C^2-1}u^{-1}(x)\,.
 \ee
 It is not difficult to check by the direct calculation that
\be
\int_{-\infty}^{\infty}\phi_{-1/2}^2(x)dx=1\,.
\ee

Using Theorem 1 and equations (\ref{G-osc}) and (\ref{flr})
we obtain the propagator for the Hamiltonian with potential
(\ref{pot1})
\begin{align}
K_c(x,y,t)=
-\frac{\sqrt{\pi}(C+1)}{\sqrt{2}u(y)}L_x\int_{-
\infty}^{y}K_0(x,z,t)e^{\frac{z^2}{4}}(1+\erf(z/\sqrt{2}))dz
\nonumber \\
+\frac{\sqrt{\pi}(C-
1)}{\sqrt{2}u(y)}L_x\int_{y}^{\infty}K_0(x,z,t)e^{\frac{z^2}{4}}(1-
\erf(z/\sqrt{2}))dz+ \phi_{-1/2}(x)\phi_{-1/2}(y)e^{it/2}\,.
\end{align}

As the second example we consider a complex extension of the
one-soliton potential
\be\label{cs}
V_c(x)=\frac{-2a^2}{\cosh^2(ax+c)}\,.
\ee
which is obtained from $V_0(x)=0$
with the choice
\[
u(x)=\cosh(ax+c)\,,\quad
\a=-a^2\,,\quad {\rm Im}\,c\ne 0\,,\quad {\rm Im}\,a=0
\]
as the transformation function.
The Hamiltonian $h_c$ with potential (\ref{cs}) has a single
discrete level $E_0=-a^2$ with the eigenfunction
\[
\phi_{-a^2}(x)=\sqrt{\frac{a}{2}}\,\frac{1}{u(x)}\,.
\]
Just like in the previous example this function is such that
\be
\int_{-\infty}^\infty\phi_{-a^2}^2(x)dx=1\,.
\ee
To find the propagator $K_c$ we are using the known propagator
and Green's function for the free particle which are
\[
K_0(x,y;t)=\frac{1}{\sqrt{4\pi it}}\, {\rm{e}}^{\displaystyle
\frac{i(x-y)^2}{4t}}\,,\quad
G_0(x,y,E)=\frac{i}{2\k}e^{i\k|x-y|}\,,\qquad
\mbox{Im}\k>0\,,\qquad E=\k^2\,.
\]
After some transformations similar to that
 described in details in \cite{PS}
we get the propagator for the Hamiltonian $h_c$
\be\label{ospp}
K_c(x,y;t)=\frac{1}{\sqrt{4\pi it}}\, {\rm{e}}^{
\frac{i(x-y)^2}{4t}}+
\frac{a{\rm{e}}^{ia^2t}}{4u(x)u(y)}\left[\erf_{+}
+\erf_{-}\right]
\ee
where the notation
$\erf_{\pm}=\erf [a\sqrt{it}\pm\frac{1}{2\sqrt{it}}(x-y)]$ is used
and for $c$ we take the value
$c={\rm arctanh}\,\frac{b^2-a^2}{2iab}$, ${\rm Im}(b)=0$.

We have to note that formula (\ref{ospp}) is in the prefect
agreement with the result obtained by Jauslin \cite{J} where the
replacement $t\to it$ should be made since this author got the
propagator for the heat equation with the  one-soliton potential.
Another
comment we would like to make is that the Jauslin's method
 is based on an integral
formula which relates solutions of two Schr\"dinger equations whose
Hamiltonians are SUSY partners.
Unfortunately,
integrals which need to be calculated
when the method is
applied to the Schr\"odinger equation  become
divergent.
Therefore the author found the propagator for the
 heat equation with the one-soliton potential.
 The Schr\"odinger equation may be considered as
 the heat equation with the imaginary time.
 In this respect the
 following question arises: whether the Jauslin's  result
 after the replacement $t\to it$
 gives the propagator for the \Sc equation with the one-soliton potential?
 We want to
 stress that the answer to this question is not trivial since such
 a replacement at the level of the Jauslin's integral transformation
 leads to divergent integrals and only the replacement in the
 final result gives the finite value for the propagator.
 Thus, our analysis shows that the answer
 to this question is positive.

In conclusion we note that in this letter we presented a method
for finding propagators for those non-Hermitian Hamiltonians
with a purely real spectrum,
SUSY partners of which have the known both the propagator and
Green's function. The general theorem is illustrated by considering
 a complex anharmonic oscillator Hamiltonian and a complex
 extension of the one-soliton potential.

 The work is partially supported by
the grants SS-5103.2006.2 and RFBR-06-02-16719
and the Russian 'Dynasty' foundation (AMP).
The authors would like to acknowledge that the
form (\ref{Ut}) of the evolution operator and its relation with the
propagator was suggested by the anonymous referee.

\end{document}